\documentclass[a4paper,10pt,twocolumn]{article}
\usepackage{ol2_modified}

\usepackage{CJK}
\usepackage{graphicx}
\usepackage{amsmath}
\usepackage{amssymb}

\newcommand{\degrees}{\ensuremath{^\circ}}
\newcommand{\half}{\ensuremath{\tfrac{1}{2}}}

\newcommand{\mat}[1]{\hat{#1}}

\newcommand{\reffig}[1]{Fig.~\ref{fig:#1}}

\newcommand{\Attenuation}{\alpha}
\newcommand{\PhaseDelay}{\beta}
\newcommand{\dB}{\ensuremath{\,\text{dB}}}

\begin{document}
\twocolumn[ 

\begin{CJK*}{GB}{} 

\title{H\"ansch--Couillaud locking of Mach--Zehnder interferometer for carrier removal from a phase-modulated optical spectrum}
\vspace{-2mm}
\author{J.E.~Bateman, R.L.D.~Murray, M.~Himsworth, H.~Ohadi, A.~Xuereb, and T.~Freegarde}
\address{School of Physics and Astronomy, University of Southampton\\Southampton SO17 1BJ, UK}

\vspace{-2mm}
\begin{abstract}%
We describe and analyse the operation and stabilization of a Mach--Zehnder interferometer, which separates the carrier and the first-order sidebands of a phase-modulated laser field, and which is locked using the H\"ansch--Couillaud method. In addition to the necessary attenuation, our interferometer introduces, via total internal reflection, a significant polarization-dependent phase delay.  We employ a general treatment to describe an interferometer with an object which affects the field along one path, and we examine how this phase delay affects the error signal.  We discuss the requirements necessary to ensure the lock point remains unchanged when phase modulation is introduced, and we demonstrate and characterize this locking experimentally.  Finally, we suggest an extension to this locking strategy using heterodyne detection.
\end{abstract}

 
\maketitle

\end{CJK*}
] 

\section{Introduction}
For many experiments and laser-locking schemes it is necessary to create light which is phase-coherent with and frequency shifted relative to a master laser oscillator.  Common approaches include acousto-optical modulation~\cite{Bouyer1996}, electro-optical modulation~\cite{Szymaniec1997}, and current modulation of the laser~\cite{Lau1984,Ringot1999,Affolderbach2000}.
 The latter two of these approaches give, typically, a phase-modulated spectrum.

It is often useful to separate the frequency components of such a spectrum, and there are several devices which can perform this task~\cite{haubrich00,Abel2009}.  We describe one of these---a Mach--Zehnder interferometer---and a locking-scheme based on the method of H\"ansch and Couillaud~\cite{hansch80}; we include a description of a relative phase delay between the linear polarization components, which is expected for many real devices, and we take care to ensure the lock point does not change when phase modulation is introduced.  This property is essential in our application, where we use a sideband from an electro-optically modulated laser field to drive Raman transitions between hyperfine states in cold alkali-earth atoms (similar to ref.~\cite{Kasevich1992}), and in which we change the modulation frequency and depth during the experiment.  For related uses, see references \cite{AP33} and \cite{schneider2009}.  We will provide a simple and general analysis, and a demonstration of a robust embodiment in a realistic experimental setting.

\section{Theoretical framework}
\begin{figure}
        \centerline{\includegraphics[width=0.6\linewidth]{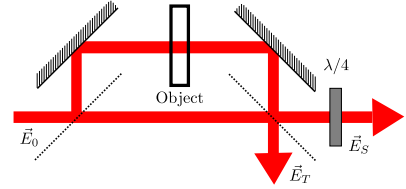}}
        \caption{\label{fig:aside}Prototype interferometer showing incident field $\vec{E}_0$, beam-splitters, output fields $\vec{E}_T$ and $\vec{E}_S$ (the latter after a quarter-wave plate $\lambda/4$) and an internal object which affects the field which follows the longer path.}
\end{figure}

Consider a light field $\vec{E}_0$ incident on an interferometer, as depicted in \reffig{aside}. One path passes unperturbed to the output beam-splitter, while the second path is subject to a phase delay $\phi$, and passes through an object, described by $\mat{O}$, before recombining with the first path. The output field is hence
\begin{equation}
 \vec{E}_T = \frac{1}{2}\left(\vec{E}_0 - e^{i\phi}\mat{O}\vec{E}_0\right)\,;
\end{equation}the minus sign is a consequence of the phase change under reflection.

We pass this field through a quarter-wave plate (with fast-axis at $45\degrees$ to the vertical) and analyze the resulting electric field:
\begin{equation}
 \vec{E}_S = \mat{Q}\,\frac{1}{2}\left(\vec{E}_0 + e^{i\phi}\mat{O}\vec{E}_0\right)\text{, where }\mat{Q} = \frac{1}{\sqrt{2}}\left(\begin{array}{cc}1&i\\i&1\end{array}\right)
\end{equation}
in the standard Jones matrix representation~\cite{hecht01}.  We define the error signal $S$ as the difference between the linear polarization components of the field $\vec{E}_S$, and the transmission $T$ as the horizontal component of the field $\vec{E}_T$:
\begin{equation}
 S = \big|\vec{E}_S\cdot\hat{x}\big|^2-\big|\vec{E}_S\cdot\hat{y}\big|^2~\text{and}\,~T = \big|\vec{E}_T\cdot\hat{x}\big|^2\,;
\end{equation}
we have discarded the vertical polarization from the output, which reduces the power but means it is possible to achieve complete extinction. 

By choosing the incident field and the internal object this framework can be used to describe a variety of locking schemes.  The original H\"ansch--Couillaud scheme, although here phrased in terms of interferometers rather than a cavity, corresponds to horizontally polarized light and a rotated linear polarizer~\cite{hansch80}.

We now apply this framework to describe an internal object which attenuates and also delays one polarization.  The linearly polarized incident field $\vec{E}_0$, and the  object $\mat{O}$ are described by the following:
\begin{equation}
 \vec{E}_0 = E_0 \left(\begin{array}{cc}\cos\theta\\\sin\theta\end{array}\right)~\text{and}~\mat{O}=\left(\begin{array}{cc}1&0\\0&(1-\Attenuation)e^{i\PhaseDelay}\end{array}\right)
\end{equation}where $\Attenuation$ is the attenuation, $\PhaseDelay$ is the phase delay, and $\theta$ is the angle by which the linearly polarized light is inclined from vertical.

In the absence of attenuation, the two polarizations will be subjected only to different delays, and the difference between the two displaced interference patterns could be used as an error signal; here the field before the quarter-wave plate would be used to derive $S$.

If the phase delay $\PhaseDelay$ were zero, we would recover a rotated version of the H\"ansch--Couillaud scheme.  Also, if the attenuation was total ($\Attenuation=1$), any phase delay would be inconsequential and we would again recover this original scheme.

In the intermediate case of partial attenuation $(\Attenuation<1)$ and non-zero phase delay $(\PhaseDelay\neq0)$, we find that the error signal crosses zero at the \emph{maximum} of transmission, but has the non-zero value $\half(1-\Attenuation)\sin\PhaseDelay\sin{2\theta}$ at the transmission minimum.
Its gradient about the extrema is $\pm\frac{1}{4}\big[1-(1-\Attenuation)\cos\PhaseDelay\big]\sin{2\theta}$.
In this intermediate regime, the introduction of phase modulation, as discussed in the next section, affects the positions at which the error signal crosses zero.  As described later, we found that a phase delay was unavoidable in our device; therefore, in order to recover the H\"ansch--Couillaud scheme, we introduced total attenuation of the vertical polarization component.

\section{Phase modulated light}
We may analyze a modulated field $\vec{E}(t)=\vec{E}_0 e^{i\int_0^t\omega(t') dt'}$ in terms of its Fourier components. For the case of simple phase modulation, where $\int_0^t\omega(t') dt'= \omega_0 t + m\cos{\Omega t}$ ($\omega_0$: unmodulated frequency; $\Omega$: modulation frequency; $m$: modulation depth), we expand using the Jacobi--Anger identity, treat the sidebands as independent fields, and sum their contributions to obtain the error signal
\begin{equation}
S_\text{PM}\big(\omega_0\tau\big) = \sum_{n=-\infty}^{+\infty}\left|J_n(m)\right|^2 S\big(\omega_0\tau+n\Omega\tau\big)\,,
\end{equation} where the subscript `PM' indicates that phase modulation is present; $\tau=\phi/\omega_0$ is the optical path delay in our device, $J_n(m)$ is the $n^\text{th}$ order Bessel function, and we have assumed that all frequency components interact in the same way with the optical elements.

We operate the device near the condition $\Omega\tau=\pi$ which separates the carrier from the first-order sidebands or, more generally, ensures odd and even numbered sidebands exit from opposite ports of the interferometer.
Unless the signal crosses zero at the transmission minimum, the position where it does cross will shift when phase modulation is introduced.
Hence we must ensure the signal is zero at the transmission extrema while maintaining a non-zero gradient; this is satisfied for $\PhaseDelay=0$ or $\Attenuation=1$.  To achieve this, we can compensate for any differential phase shift $\PhaseDelay$ in our device using a waveplate, or we can introduce complete attenuation of the vertical polarization using a linear polarizer.
The overall signal becomes
\begin{equation}
S_\text{PM}(\omega_0\tau)=S\big(\omega_0\tau\big)\sum_{n=-\infty}^{+\infty} (-1)^n \left|J_n(m)\right|^2\label{eqn:PMresult} .
\end{equation}

Under these conditions, any change in phase modulation depth does not affect the positions where the error signal crosses zero; the gradient changes, but for modulation depths $m\lesssim0.38\pi$ it maintains the same sign.

\section{Experimental implementation}
A Mach--Zehnder interferometer was constructed from readily available components, including two non-polarizing BK7 beam-splitter cubes, and a BK7 right-angled prism\cite{thorlabs}.  The cubes were glued using low-expansion UV curing glue\cite{dymax}, with care taken to ensure their faces were parallel, and the pair was mounted on a kinematic mount.  The prism was glued to a translation stage; a screw was available for coarse path-difference adjustment, and a piezo-electric stack was used for small adjustments and locking feedback.  Incident light was spatially filtered and collimated (to ensure the incident wavefronts were flat) and aligned into the device.  Beam overlap was found visually, and then maximized by scanning the path difference using the piezo and observing the contrast ratio of power exiting each of the output ports.
\begin{figure}
        \centerline{\includegraphics[angle=-90,width=\linewidth]{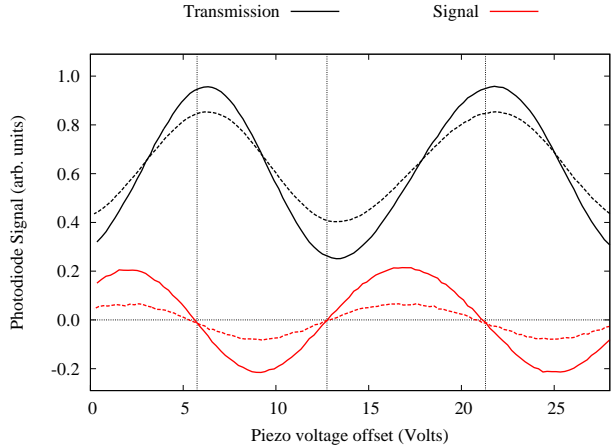}}
        \caption{\label{fig:throughput}Transmission $T$ (black) and error signal $S$ (red) with (solid) and without (dashed) phase modulation of the input light, as the path difference is scanned using a piezo-electric stack.  The dashed vertical lines mark the positions where the error signals coincide, and these are very close to $S=0$.  They are, however, displaced from the transmission extrema, but there is adequate scope for optimization for a given input field.}
\end{figure}

In the absence of a relative phase delay between the polarizations, one could generate an error signal using the slight polarization selectivity of reflections by the nominally polarization insensitive beam-splitter cubes.
However, we found that a significant relative phase delay of $\PhaseDelay\approx78\degrees$ was introduced by the two total internal reflections inside our right-angled prism~\cite{hecht01} and, to recover the H\"ansch--Couillaud method, it was necessary to introduce a linear polarizer to ensure complete attenuation of the vertical polarization (i.e. $\Attenuation=1$).

\begin{figure}
        \centerline{\includegraphics[angle=-90,width=\linewidth]{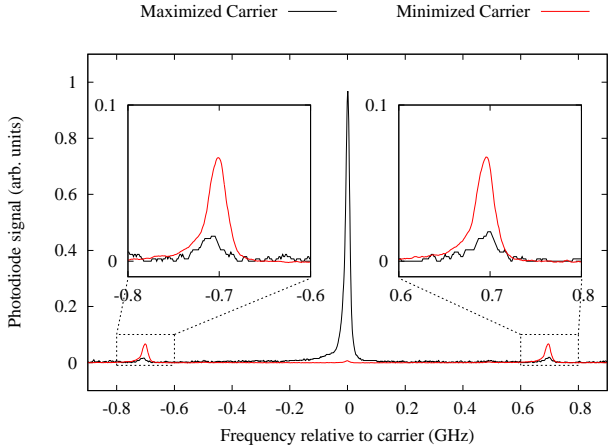}}
        \caption{\label{fig:spectra}Phase-modulated spectra after filtering by the Mach--Zehnder interferometer, showing maximum and minimum carrier transmission.  The modulation frequency ($2.7\,\text{GHz}$) is larger than the free-spectral range of the scanning Fabry--P\'erot cavity ($2\,\text{GHz}$) used to obtain this trace, and so the sidebands appear at $\pm700\,\text{MHz}$ for the lower- and upper-sidebands, respectively; these are magnified in the insets.  By comparing the amplitudes, we estimate the modulation depth to be $m\approx0.2\pi$, and by comparing with an unfiltered reference trace (obtained by blocking one path of the interferometer), we see that, for the case of minimized carrier transmission, the carrier is attenuated by more than $20\dB$ while approximately $1\dB$ of sideband power is lost. This trace was smoothed using a $5\,\text{MHz}$ bandwidth moving-average filter.}
\end{figure}

\reffig{throughput} shows the interferometer transmission measured with and without the phase-modulated optical sidebands that we wish to separate. The modulation frequency is  $\Omega=2\pi\times2.725\,\text{GHz}$ and the wavelength is $\lambda=780\,\text{nm}$; 
the device is operated near to a path difference $c\tau=c\pi/\Omega\approx55\,\text{mm}$, which corresponds to the condition $\Omega\tau=\pi$.
The photodiodes sample different parts of the beam cross-section, and also have different responsivities.  It may be necessary to adjust the photodiode balance and the offset, but as demonstrated by this unoptimized trace, a real device operates approximately as predicted.  From the decrease in visibility and error signal, we estimate a modulation depth of $m\approx0.2\pi$.  This agrees with the more direct measurement with a scanning Fabry--P\'erot cavity, as shown in \reffig{spectra}.  The imperfect behaviour of the device is accounted for partly by the unequal reflectivities of the beam-splitter cubes, but a more significant problem is the spatial overlap of the fields in this free-space device; we would expect improved performance and more ideal behaviour if the device was implemented using single-mode optical fibers and fiber-based beam-splitters.

\begin{figure}
        \centerline{\includegraphics[angle=-90,width=\linewidth]{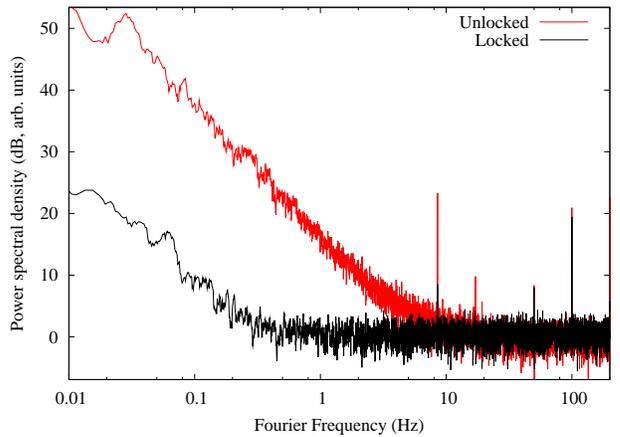}}
        \caption{\label{fig:errorFFT}Fast Fourier transform of the recorded output when the system is locked (black) and unlocked (red).  The plot has been smoothed using a $10\,\text{mHz}$ bandwidth moving-average filter.}
\end{figure}

We constructed a feedback circuit with an integrated high-voltage output, similar to that in ref.~\cite{yashchuk00}, and recorded the transmission with and without feedback; the Fourier transforms of these are shown in \reffig{errorFFT}. The bandwidth of the circuit is $\sim100\,\text{Hz}$, and for low frequencies ($<10\,\text{Hz}$) the circuit reduces drift by several orders of magnitude.  The very slightly increased noise at high frequency is expected for a feedback circuit.

\section{Alternative: Heterodyne detection}
Another method by which we could ensure the error signal would remain unchanged when phase modulation was introduced would be to mix the photodiode signals with a frequency-shifted reference, derived from the unmodulated field, and extract electronically a signal corresponding to the beat note of this reference mixed with the carrier.
Furthermore, it would be possible to tune the reference field frequency close to, and hence make the error signal depend upon, any chosen frequency component in a complex, more powerful spectrum.  Alternatively, with a sufficiently fast photodiode, one may tune the detection electronics to select a desired optical frequency component.

\begin{figure}
	\centerline{\includegraphics[width=0.8\linewidth]{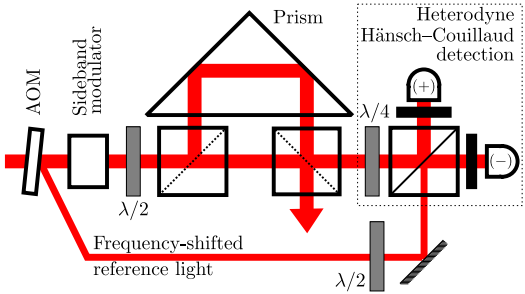}}
        \caption{\label{fig:heterodyne}A proposed H\"ansch--Couillaud scheme using heterodyne detection.  An acousto-optical modulator (AOM) extracts a reference before sideband modulation. Light passes through the interferometer, composed of non-polarizing beam-splitter cubes and a right-angled prism, and is detected by photodiodes labeled $(\pm)$. The reference light is incident on the detectors and the linear polarizers (solid black lines) are at $45\degrees$ relative to the beam-splitter cube. The two half-wave plates labeled $\lambda/2$ serve to introduce the small rotation necessary for the H\"ansch--Couillaud scheme before the Mach--Zehnder, and to rotate the reference light by $45\degrees$ so that equal power from this reference falls on the photodiodes after the polarizing beamsplitter cube in the heterodyne detection setup.}
\end{figure}

This reference field could be created by using an acousto-optical modulator (AOM) to pick off a small fraction of the laser field before it passes through, for example, an electro-optical phase modulator; see \reffig{heterodyne}.  As an example, consider an interferometer which separates positive and negative first-order sidebands from a $3\textrm{GHz}$ phase-modulated field.  Using a $100\textrm{MHz}$ AOM-shifted reference field and a fast photodiode, one could detect beat-frequencies at $2900\text{MHz}$, $100\textrm{MHz}$, and $3100\textrm{MHz}$; using standard electronics, a specific frequency component could be extracted and passed to the locking circuit.  Here, feedback could be made to depend on one sideband, leaving the other sideband and a significant fraction of the carrier. The resulting spectrum would be well suited to driving stimulated Raman transitions.

\section{Conclusions}
We have described and demonstrated a Mach--Zehnder interferometer used to separate carrier and first-order sidebands from a phase-modulated laser field, which we have locked using the H\"ansch--Couillaud method.  The aim of this article was twofold. Firstly, we provided a simple model that allows the interferometer to be understood in terms of its constituent birefringent and reflective elements. H\"ansch and Couillaud did not consider a differential phase-change between the polarisations, but this arises naturally in our real device and so we have extended their model. We then applied this model to the analysis of the interferometer used in our experiments; our results show that the technique is, despite its simplicity, appropriate for this commonly-encountered situation.

The polarization-dependent phase delay, originating from total internal reflection in the corner-reflector of the Mach--Zehnder interferometer, affects the error signal, causing an offset at the transmission minimum and leaving the locking scheme sensitive to intensity fluctuations.  These effects can be eliminated by extinguising one polarization using an internal linear polarizer, leaving the lock point approximately fixed as phase modulation is introduced. The slight residual offset of the lock points from the transmission extrema observed experimentally was readily corrected by adjusting the photodiode balance, and the interferometer was easily optimized for a given input spectrum. An alternative approach accomplishes this indifference to phase modulation using heterodyne detection and a frequency-shifted reference light field.

\bibliographystyle{osajnl}

\begin{thebibliography}{10}
\newcommand{\enquote}[1]{``#1''}

\bibitem{Bouyer1996}
P.~Bouyer, T.~L. Gustavson, K.~G. Haritos, and M.~A. Kasevich,
  \enquote{{Microwave signal generation with optical injection locking},}
  Optics Letters \textbf{21}, 1502 (1996).

\bibitem{Szymaniec1997}
K.~Szymaniec, \enquote{{Injection locking of diode lasers to frequency
  modulated source},} Optics Communications \textbf{144}, 50--54 (1997).

\bibitem{Lau1984}
K.~Y. Lau, C.~Harder, and A.~Yariv, \enquote{{Direct modulation of
  semiconductor lasers at $f>10\,$GHz by low-temperature operation},} Applied
  Physics Letters \textbf{44}, 273 (1984).

\bibitem{Ringot1999}
J.~Ringot, Y.~Lecoq, J.~Garreau, and P.~Szriftgiser, \enquote{{Generation of
  phase-coherent laser beams for Raman spectroscopy and cooling by direct
  current modulation of a diode laser},} The European Physical Journal D
  \textbf{7}, 285 (1999).

\bibitem{Affolderbach2000}
C.~Affolderbach, A.~Nagel, S.~Knappe, C.~Jung, D.~Wiedenmann, and R.~Wynands,
  \enquote{{Nonlinear spectroscopy with a vertical-cavity surface-emitting
  laser (VCSEL)},} Applied Physics B: Lasers and Optics \textbf{70}, 407--413
  (2000).

\bibitem{haubrich00}
D.~Haubrich, M.~Dornseifer, and R.~Wynands, \enquote{Lossless beam combiners
  for nearly equal laser frequencies,} Rev. Sci. Inst. \textbf{71}, 338--340
  (2000).

\bibitem{Abel2009}
R.~P. Abel, U.~Krohn, P.~Siddons, I.~G. Hughes, and C.~S. Adams,
  \enquote{{Faraday dichroic beam splitter for Raman light using an
  isotopically pure alkali-metal-vapor cell},} Optics Letters \textbf{34}, 3071
  (2009).

\bibitem{hansch80}
T.~W. H\"ansch and B.~{Couillaud}, \enquote{{Laser frequency stabilization by
  polarization spectroscopy of a reflecting reference cavity},} Optics
  Communications \textbf{35}, 441--444 (1980).

\bibitem{Kasevich1992}
M.~Kasevich and S.~Chu, \enquote{{Measurement of the gravitational acceleration
  of an atom with a light-pulse atom interferometer},} Applied Physics B
  Photophysics and Laser Chemistry \textbf{54}, 321--332 (1992).

\bibitem{AP33}
I.~Dotsenko, W.~Alt, S.~Kuhr, D.~Schrader, M.~Muller, Y.~Miroshnychenko,
  V.~Gomer, A.~Rauschenbeutel, and D.~Meschede, \enquote{Application of
  electro-optically generated light fields for raman spectroscopy of trapped
  cesium atoms,} Applied Physics B \textbf{78}, 711--717 (2004).

\bibitem{schneider2009}
J.~Schneider, O.~Gl\"{o}ckl, G.~Leuchs, and U.~L.~Andersen,
  \enquote{{Quadrature measurements of a bright squeezed state via sideband swapping},}
  Optics Letters \textbf{34}, 1186 (2009).

\bibitem{hecht01}
E.~Hecht, \emph{Optics} (Addison Wesley, 2001), 4th ed.

\bibitem{thorlabs}
 {Suitable components are available from many suppliers, including Thorlabs (BS011 and PS911).}

\bibitem{dymax}
 {A suitable low-expansion glue is manufactured by Dymax; part number OP-67-LS.}

\bibitem{yashchuk00}
V.~V. Yashchuk, D.~Budker, and J.~R. Davis, \enquote{Laser frequency
  stabilization using linear magneto-optics,} Rev. Sci. Inst. \textbf{71},
  341--346 (2000).

\end{thebibliography}

\end{document}